\documentclass[aps,pra,superscriptaddress,reprint,floatfix,longbibliography,showpacs]{revtex4-1}
\usepackage[english]{babel}
\usepackage{amsmath}
\usepackage{amssymb}
\usepackage{graphicx}
\usepackage{hyperref}
\usepackage{upgreek}
\usepackage{dcolumn}

\usepackage{xspace}
\usepackage{xcolor}

\def\lirho{Li$_2$RhO$_3$}
\def\liiro{$\alpha$-Li$_2$IrO$_3$}

\def\rucl{$\alpha$-RuCl$_3$}

\def\p1bar{$P{\overline1}$}

\begin{document}
\title{Pressure-induced formation of rhodium zigzag chains \\ in the honeycomb rhodate Li$_2$RhO$_3$}

\author{V.~Hermann}
\affiliation{Experimentalphysik II, Augsburg University, 86159 Augsburg, Germany}
\author{S.~Biswas}
\affiliation{Institut f\"{u}r Theoretische Physik, Goethe-Universit\"{a}t Frankfurt, 60438 Frankfurt am Main, Germany}
\author{J.~Ebad-Allah}
\affiliation{Experimentalphysik II, Augsburg University, 86159 Augsburg, Germany}
\affiliation{Department of Physics, Tanta University, 31527 Tanta, Egypt}
\author{F.~Freund}
\author{A.~Jesche}
\author{A.~A.~Tsirlin}
\affiliation{Experimentalphysik VI, Center for Electronic Correlations and Magnetism, Augsburg University, 86159 Augsburg, Germany}
\author{M.~Hanfland}
\affiliation{European Synchrotron Radiation Facility (ESRF), BP 220, 38043 Grenoble, France}
\author{D.~Khomskii}
\affiliation{II. Physikalisches Institut, Universit\"{a}t zu K\"{o}ln, Z\"{u}lpicher Strasse, 77, 50937, K\"{o}ln, Germany}
\author{P.~Gegenwart}
\affiliation{Experimentalphysik VI, Center for Electronic Correlations and Magnetism, Augsburg University, 86159 Augsburg, Germany}
\author{R.~Valent\'{i}}
\affiliation{Institut f\"{u}r Theoretische Physik, Goethe-Universit\"{a}t Frankfurt, 60438 Frankfurt am Main, Germany}
\author{C.~A.~Kuntscher}
\affiliation{Experimentalphysik II, Augsburg University, 86159 Augsburg, Germany}
\email{christine.kuntscher@physik.uni-augsburg.de}

\begin{abstract}
We use powder x-ray diffraction to study the effect of pressure on the crystal structure of the honeycomb rhodate Li$_2$RhO$_3$. We observe low-pressure ($P$$<$$P_{c1}$ = 6.5 GPa) and high-pressure ($P$$>$$P_{c2}$ = 14 GPa) regions corresponding to the monoclinic $C2/m$ symmetry, while a phase mixture is observed at intermediate pressures. At $P$$>$$P_{c2}$, the honeycomb structure becomes distorted and features short Rh--Rh bonds forming zigzag chains stretched along the crystallographic $a$ direction. This is in contrast to dimerized patterns observed in triclinic high-pressure polymorphs of $\alpha$-Li$_2$IrO$_3$ and $\alpha$-RuCl$_3$. Density-functional theory calculations at various pressure conditions reveal that the observed rhodium zigzag-chain pattern is not expected under hydrostatic pressure but can be reproduced by assuming anisotropic pressure conditions.
\end{abstract}
\pacs{}

\maketitle

\section{Introduction}

In recent years, $4d$ and $5d$ transition-metal compounds were intensively studied due to their extremely rich physics. In comparison to $3d$ compounds, where the electronic correlation $U$ dominates over the spin-orbit coupling constant $\lambda\textsubscript{SOC}$ and Hund's coupling $J\textsubscript{H}$, spin-orbit coupling (SOC) becomes more and more important for $4d$ and $5d$ transition-metal compounds, whereas the strength of electronic correlations decreases. The actual physics of these compounds thereby depends on a delicate balance between $U$, $\lambda\textsubscript{SOC}$ and $J\textsubscript{H}$, as well as the crystal structure. The class of layered honeycomb-type $4d$ and $5d$ transition-metal compounds, such as $A_2M$O$_3$ ($A$= Li, Na and $M$= Ir, Rh) and $\alpha$-RuCl$_3$, is especially interesting in this regard, as this delicate balance of parameters was discussed in terms of Kitaev physics and possible spin-liquid state \cite{Kitaev.2006, Jackeli.2009, Chaloupka.2010, Choi.2012, Plumb.2014, Chun.2015, Winter.2017,Luo.2013,Khuntia.2017}. However, in Na$_2$IrO$_3$, $\alpha$-Li$_2$IrO$_3$, and $\alpha$-RuCl$_3$ the quantum spin liquid ground state is not realized, since these materials were found to order magnetically at low temperatures~\cite{Choi.2012, Sears.2015, Banerjee.2016, Williams.2016}.

As for Li$_2$RhO$_3$, its magnetic ground state is still under debate. No long-range magnetic order could be found down to $\approx$0.5~K, but instead at small magnetic fields spin freezing was observed below 6-7~K~\cite{Luo.2013,Khuntia.2017}, although it is suspected that the majority of magnetic moments form a fluctuating liquid-like state~\cite{Khuntia.2017}. Whether this partial spin freezing is due to a proximity to the Kitaev quantum spin liquid ground state or due to unavoidable defects (anti-site disorder and/or stacking faults) is still unclear~\cite{Khuntia.2017,Katakuri.2015}. However, \textit{ab initio} and effective-model calculations showed that Li$_2$RhO$_3$ bears similar electronic structure to the iridates~\cite{Mazin.2013} and hosts anisotropic Kitaev interaction terms of the same magnitude as in $5d$ iridates~\cite{Katakuri.2015}. According to electrical resistivity measurements, Li$_2$RhO$_3$ is insulating at ambient pressure \cite{Luo.2013,Mazin.2013}.

Another interesting aspect of Li$_2$RhO$_3$ is its behavior under pressure, where honeycomb iridates~\cite{Hermann.2018,Clancy.2018} and $\alpha$-RuCl$_3$~\cite{Bastien.2018,Biesner.2018} become dimerized and, consequently, non-magnetic. Previously~\cite{Hermann.2018}, we showed that the size of the central ion, the strength of the spin-orbit-coupling, electronic correlations, and Hund's coupling all act against the dimerization. In comparison to $\upalpha$-Li$_2$IrO$_3$, the $\lambda\textsubscript{SOC}$ in Li$_2$RhO$_3$ is expected to be lower, while the electronic correlations should be enhanced in Rh$^{+4}$ compared to Ir$^{4+}$, as screening by oxygen orbitals is reduced. Thereby, one generally expects a higher transition pressure in Li$_2$RhO$_3$ and a larger pressure range for tuning the putative Kitaev magnetism of this compound. Here, we show that this is the case, but also that the pressure-induced transformations are very different from the dimerization observed in honeycomb iridates.

\section{Methods}

\subsection{Experimental Details}
A powder sample of Li$_2$RhO$_3$ was prepared by a solid-state reaction of Li$_2$CO$_3$ and Rh in oxygen flow at $850^{\circ}$\,C with several intermediate re-grindings. The sample quality was confirmed by laboratory x-ray diffraction (XRD) using the Rigaku MiniFlex diffractometer (CuK$_{\alpha}$ radiation, Bragg-Brentano geometry). This synthesis procedure yields samples with the best structural order achieved so far~\cite{Mannithesis}, although stacking faults are still present. Their concentration is discussed in Sec.~\ref{sec:exp} below.

Li$_2$RhO$_3$ powder was loaded into a diamond anvil cell (DAC) for pressure generation, and helium was used as pressure transmitting medium. The powder x-ray diffraction patterns were obtained using synchrotron radiation at the beamline ID15B at the European Synchrotron Radiation Facility (ESRF), Grenoble at room temperature. The wavelength of the radiation was 0.411267~{\AA}, and the patterns were obtained in the 2$\theta$-range between 2 and 33$^\circ$. The pressure in the DAC was determined \textit{in situ} by the ruby luminescence method. The resulting patterns were analyzed by Rietveld refinements using the Jana2006 software \cite{Petricek.2014}. The quality of the fit is gauged by the weighted structure factor $R_\omega$, as defined in Ref.~\onlinecite{Young.2002} and by the commonly used weighted profile factor $R_{\omega  p}=(\sum_{i}w_i(y^\prime_i(\text{\rm obs})-y^\prime_i(\text{\rm calc}))^2)/(\sum_{i}w_i y^\prime_i(\text{\rm obs})^2)$, where $y^\prime_i$ are intensities corrected for background. The absorption correction for a cylindrical sample was calculated to be lower than the value one (using Ref. \cite{Dreele.2013}), so no absorption correction was applied. The isotropic atomic displacement parameters $U_{\rm iso}$ were fixed to the value of 0.005~\AA$^2$ for all atomic positions, except for Rh(1) and Li(1).

\subsection{Computational details}
\label{sec:method-comput}
Structural optimizations were performed under different pressure conditions by using projector-augmented planewave \cite{blochl} method based on density functional theory (DFT), as implemented in the VASP package \cite{Kresse}. Calculations were done within the generalized gradient approximation (GGA),  GGA+U \cite{Dudarev}, and GGA+SOC+U (including spin-orbit coupling effects for Rh). The value of the on-site Coulomb parameter $U$  was chosen based on the reproducibility of the experimental structure, as will be shown below. The cutoff for the wavefunction was set at 650 eV. K-point meshes of size $8\times6\times8$ were used for all the structural optimizations.

\begin{figure}[!t]
        \includegraphics[width=0.9\columnwidth]{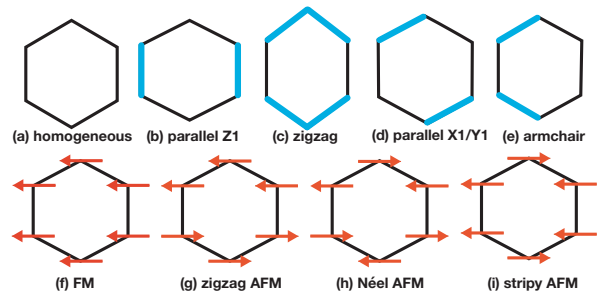}
       \caption{Schematics of  (a)-(e) various types of possible dimerization  in hexagonal Kitaev systems and  (f)-(i) different magnetic  configurations considered by us for Li$_2$RhO$_3$. The blue lines indicate the short bond, $l_s$, i.e., the dimer and the red arrows indicate the spin orientation at the transition metal site.}
         \label{fig1}
\end{figure}

We performed two types of structural optimizations: (i) allowing relaxation of both lattice parameters and atomic positions under fixed hydrostatic pressure; we refer to this as ``full relaxation", and (ii) keeping the lattice parameters fixed according to given pressure conditions and allowing only the relaxation of the atomic positions. In both cases, the system is allowed to relax until the total force acting on the system was less than 0.005 eV/\AA. At each pressure value, several different initial magnetic configurations were considered: (a) ferromagnetic (FM), (b) zigzag antiferromagnetic (AFM), (c) Ne\'{e}l AFM, (d) stripy AFM, and (e) nonmagnetic (NM) (see Fig.~\ref{fig1}).

The analysis of the electronic properties was done with the Full Potential Local Orbital (FPLO) basis \cite{fplo}.

\section{Results and Discussion}
\subsection{Experimental results}
\label{sec:exp}
\begin{figure}[t]
\includegraphics[width=0.9\columnwidth]{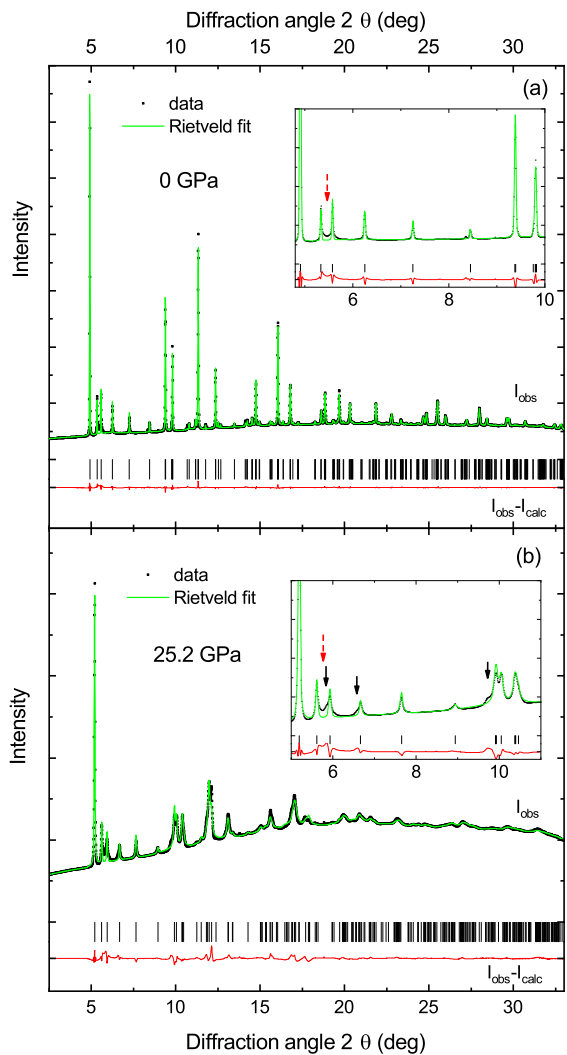}
\caption{X-ray powder diffraction diagrams ($I_{\rm obs}$) of Li$_2$RhO$_3$ at (a) the ambient pressure and (b) the highest studied pressure (25.2~GPa) together with the corresponding Rietveld fits $I_{\rm calc}$ and the difference curves ($I_{\rm obs}$-$I_{\rm calc}$). Markers indicate the calculated peak positions. The $R_\omega$ ($R_{\omega p}$) values amount to 6.31\% (13.30\%) and 6.15\% (19.33\%), respectively. The insets in (a) and (b) show the respective low-angle region at 0 and 25.2~GPa. The dashed red arrows in the insets mark the additional intensity due to stacking faults, while the black arrows in the inset of (b) mark traces from the low-pressure phase as discussed in the text.}\label{fig.powder}
\end{figure}

The x-ray powder diffraction patterns at ambient pressure and at the highest studied pressure (25.2~GPa) together with the corresponding fits from the Rietveld refinement are displayed in Fig.~\ref{fig.powder}. Both refinements were performed within the monoclinic unit cell with the $C2/m$ symmetry. The same crystal symmetry is found for the closely related honeycomb iridates~\cite{OMalley.2008,Freund.2016,Choi.2012}. In the refinements, stacking faults associated with shifts between successive LiRh$_2$ layers were taken into account, as observed in $\alpha$-Li$_2$IrO$_3$ \cite{OMalley.2008,Freund.2016} and other Li$_2M$O$_3$ ($M$ = Mn, Pt, Ru) compounds \cite{OMalley.2008, Breger.2005, CasasCabanas.2007}. Stacking faults affect the intensity and lineshape of several peaks and lead to an additional intensity between the (020) and (110) peaks as marked by the dashed red arrows in the insets of Fig.\ \ref{fig.powder}. The presence of stacking faults was taken into account by introducing the Li/Rh mixing for the Rh(1)/Li(1) and Li(2)/Rh(2) sites while constraining the overall stoichiometry to Li$_2$RhO$_3$. This reproduces the peak intensity but not their shape \cite{OMalley.2008}.

\begin{figure}[!h]
	\includegraphics[width=0.9\columnwidth]{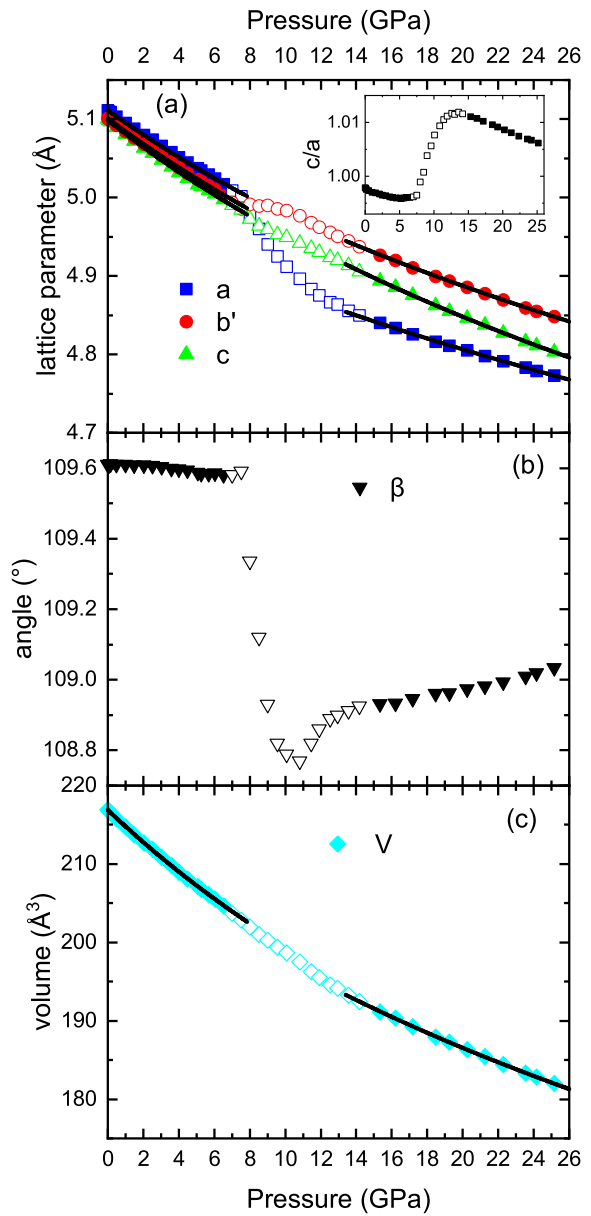}
	\caption{Pressure evolution of (a) the lattice parameters ($a,b^\prime = b/\sqrt{3},c$) and $c/a$-value (inset), (b) the monoclinic angle $\beta$ and $b/a$-value (inset, dashed line at $b/a=\sqrt{3}$) (c) the volume $V$ of the unit cell. The solid lines are fits with a Murnaghan equation of state as explained in the text. Open symbols mark the intermediate pressure regime, where the results may be less accurate due to the phase mixture (see text).}\label{fig.latticepar}
\end{figure}

The lattice parameters as a function of pressure, as obtained by the Rietveld fits of the x-ray powder diffraction diagrams, are depicted in Fig.\ \ref{fig.latticepar}.  Up to the critical pressure $P_{c1}$=6.5~GPa, the lattice parameters $a$, $b^\prime = b/\sqrt{3}$, and $c$ decrease monotonically with increasing pressure in a very similar manner. The $c/a$ value, shown in the inset of Fig.\ \ref{fig.latticepar}(a), reveals that the strongest pressure-induced effect occurs for the lattice parameter $c$. The monoclinic angle $\beta$ decreases slightly but monotonically within this pressure range.

Above $P_{c1}$, a second phase with the same $C2/m$ symmetry appears and gets more pronounced with increasing pressure. Above the critical pressure $P_{c2}$=14~GPa, this second phase is dominant and the high-pressure diffractograms can be well described by a single phase with the $C2/m$ symmetry. There are only traces of the low-pressure phase found in the diffraction patterns above $P_{c2}$ and up to the highest studied pressure, marked with black arrows in the inset of Fig. \ref{fig.powder}(b).  Most importantly, we can rule out a symmetry lowering above $P_{c2}$, as such a symmetry lowering would induce peak splittings, for example for the (021) and the (111) diffraction peaks. These peaks are observed at 7.7$^\circ$ and 9.0$^\circ$ in the inset of Fig. \ref{fig.powder}(b) and are obviously not split. Thus, both the low-pressure ($P$$<$$P_{c1}$) and high-pressure ($P$$>$$P_{c2}$) phases in Li$_2$RhO$_3$ have the $C2/m$ symmetry. This result is in contrast to the recent findings for $\alpha$-Li$_2$IrO$_3$, where a pressure-induced structural phase transition with symmetry lowering from monoclinic to the triclinic symmetry caused by the Ir--Ir dimerization occurs at 3.8~GPa  \cite{Hermann.2018}. Analogously, the monoclinic to triclinic symmetry lowering with the Ru-Ru dimerization is observed in $\alpha$-RuCl$_3$ at $P\approx1$~GPa~\cite{Bastien.2018}.

The refinement of the diffraction patterns for the intermediate pressure range, $P_{c1}$$<$$P$$<$$P_{c2}$, with a phase mixture of the low-pressure and high-pressure phases did not yield stable fits, as many of the peaks of the two phases are broad and overlapping.  Since the refinement with only one phase does not reproduce the actual peak shape, we marked this range with open symbols in Figs.\
\ref{fig.latticepar}, \ref{fig.Rh-Rh_dist}, \ref{fig.Rh-O} and
\ref{fig.oct-distortion}.

Between $P_{c1}$ and $P_{c2}$, the lattice parameter $a$ decreases drastically by about 3\,\%, while there is only a slight but abrupt increase in the lattice parameter $b$, and $c$ follows the pressure-induced monotonic decrease as observed below $P_{c1}$ [see Fig.\ \ref{fig.latticepar}(a)]. The abrupt decrease in the $a$ parameter is also revealed by the abrupt increase in the $c$/$a$ ratio. Accordingly, the most pronounced pressure-induced change happens along the $a$ lattice direction, as will be discussed in more detail later. The monoclinic angle $\beta$ abruptly decreases above $P_{c1}$, and above $P_{c2}$ it monotonically increases with increasing pressure [see Fig.\ \ref{fig.latticepar}(b)]. The kink in the pressure evolution of $\beta$ in the pressure range 10-12~GPa, i.e., in the intermediate phase, is not discussed here, because the phase mixture affects the refinements in this pressure range.

\begin{table}[b]
	\caption{Bulk moduli $B_{0,V}$ and $B_{0,r}$ with $r=a,b,c$ in the low-pressure ($P$$<$6.5~GPa) and high-pressure ($P$$>$14~GPa) phases, as obtained from fitting the volume $V$ and lattice parameters $r$ with a MOS, with $B_0^\prime$ set to 4.}\label{tab.MOS}
	\begin{ruledtabular}
		\begin{tabular}{ccc}
			& $P<6.5$~GPa & $P>15$~GPa\\
			$V_0$ (\AA$^3$) & 216.90(3) & 212.21(16)\\
			$B_{0,V}$ (GPa) & 100.4(4) & 118.6(9)\\
			$B_{0,a}$ (GPa) & 105.7(5) & 155.6(15)\\
			$B_{0,b}$ (GPa) & 100.2(6) & 122.2(12)\\
			$B_{0,c}$ (GPa) & 94.8(7) & 93.9(8)
		\end{tabular}
	\end{ruledtabular}
\end{table}

\begin{table*}
	\caption{Structural parameters for the low-pressure phase at ambient
pressure and for the high-pressure phase at 25.2~GPa. At ambient pressure, the
lattice parameters are $a=5.11126(10)$\,\r A, $b=8.83473(16)$\,\r A, $c=5.10034(11)$\,\r A,
$\beta=109.6105(18)^{\circ}$ ,$V=216.955(8)$\,\r A$^3$, and at 25.2~GPa $a=4.7732(5)$\,\r A,
$b=8.3980(7)$\,\r A, $c=4.8027(3)$\,\r A, $\beta=109.034(11)^{\circ}$, $V=181.99(3)$\,\r A$^3$. The isotropic
atomic displacement parameters $U_{\rm iso}$ were fixed to 0.005~\AA$^2$ for
all atomic positions, except for the Rh(1)/Li(1) one.}\label{tab.atomic} \begin{ruledtabular}
		\begin{tabular}{ccccccc||ccccc}
			& & \multicolumn{5}{c||}{low-pressure phase (0~GPa)} & \multicolumn{5}{c}{high-pressure phase (25.2~GPa)}\\
			Atom & Site & $x$ & $y$ & $z$ & \text{Occupancy} & $U_\text{iso}$({\AA}$^2$) & $x$ & $y$ & $z$ & \text{Occupancy} & $U_\text{iso}$({\AA}$^2$) \\\hline
			& & & & & & & & & & & \\
			Rh(1) & $4g$ & 0 & 0.3311(2) & 0 & 0.864(3) & 0.0029(3) & 0 & 0.3225(5) & 0 & 0.899(7) & 0.0047(11) \\
			Li(1) & $4g$ & 0 & 0.3311(2) & 0 & 0.136(3) & 0.0029(3) & 0 & 0.3225(5) & 0 & 0.101(7) & 0.0047(11) \\
			Li(2) & $2a$ & 0 & 0 & 0 & 0.728(3) & 0.005 & 0 & 0 & 0 & 0.798(7) & 0.005 \\
			Rh(2) & $2a$ & 0 & 0 & 0 & 0.273(3) & 0.005 & 0 & 0 & 0 & 0.202(7) & 0.005 \\
			Li(3) & $4h$ & 0 & 0.820(3) & 0.5 & 1 & 0.005 & 0 & 0.808(8) & 0.5 & 1 & 0.005 \\
			Li(4) & $2d$ & 0 & 0.5 & 0.5 & 1 & 0.005 & 0 & 0.5 & 0.5 & 1 & 0.005 \\
			O(1) & $8j$ & 0.252(17) & 0.3209(7) & 0.7631(10) & 1 & 0.005 & 0.271(3) & 0.3332(16) & 0.754(3) & 1 & 0.005 \\
			O(2) & $4i$ & 0.274(2) & 0 & 0.7726(19) & 1 & 0.005 & 0.287(4) & 0 & 0.774(4) & 1 & 0.005
		\end{tabular}
	\end{ruledtabular}
\end{table*}

The pressure dependencies of the volume $V$ and the lattice parameters $r$ ($r=a,b,c$) were fitted separately for the low- and high-pressure phases, neglecting the intermediate regime, with a second-order Murnaghan equation of state (MOS) \cite{Murnaghan.1944}, to obtain the bulk moduli $B_{0,V}$ and $B_{0,r}$ according to:
\begin{eqnarray}
V(p) & = & V_0\left[\left(B_0^\prime/B_{0,V}\right)p+1\right]^{-1/B^\prime_0},\\
r(p) & = & r_0\left[\left(B_0^\prime/B_{0,r}\right)p+1\right]^{-1/3B^\prime_0}
\end{eqnarray}
with $B^\prime$ fixed to 4. The results are summarized in Table~\ref{tab.MOS}. The bulk modulus $B_{0,V}$ of the low- and high-pressure phases amounts to 100.4(4)~GPa and 118.6(9)~GPa, respectively. This means that the material is less compressible in the high-pressure phase.  In the low-pressure phase ($P$$<$$P_{c1}$), the contribution of the $c$ direction to the bulk modulus is the lowest with $B_{0,c}$=94.8(7)~GPa, as already indicated by the pressure dependence of the $c$/$a$ ratio [inset of Fig.\ \ref{fig.latticepar} (a)]. Thus, the material is most compressible along the $c$ direction. The largest contribution to the bulk modulus is attributed to the $a$ crystal direction, with $B_{0,a}$=105.7(5)~GPa.

In the high-pressure phase ($P>P_{c2}$), the contribution $B_{0,c}$ of the $c$ direction remains low and is even slightly decreased as compared to the low-pressure phase. Most interestingly, the contribution $B_{0,a}$ is strongly increased to 155.6(15)~GPa in the high-pressure phase, while $B_{0,b}$ is much less increased, i.e., to 122.2(12)~GPa. Hence, the honeycomb layers along the $ab$-plane become less compressible in the high-pressure phase, whereby the pressure-induced hardening has the strongest effect along the $a$ direction.

\begin{figure}[b]
	\includegraphics[width=\columnwidth]{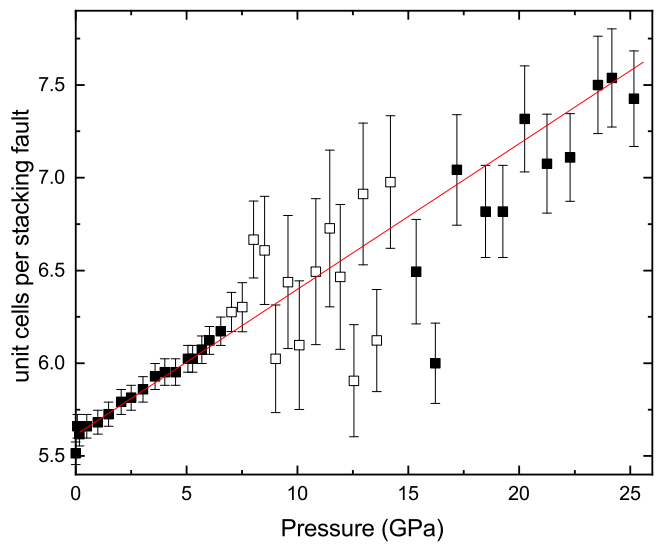}
	\caption{\label{fig.stackingfaults}Average number of unit cells per one stacking faults, as estimated from the fractional occupation of the Li(1)/Rh(1) site. The red line is a guide to the eye.}
\end{figure}

\begin{figure}
	\includegraphics[width=0.95\columnwidth]{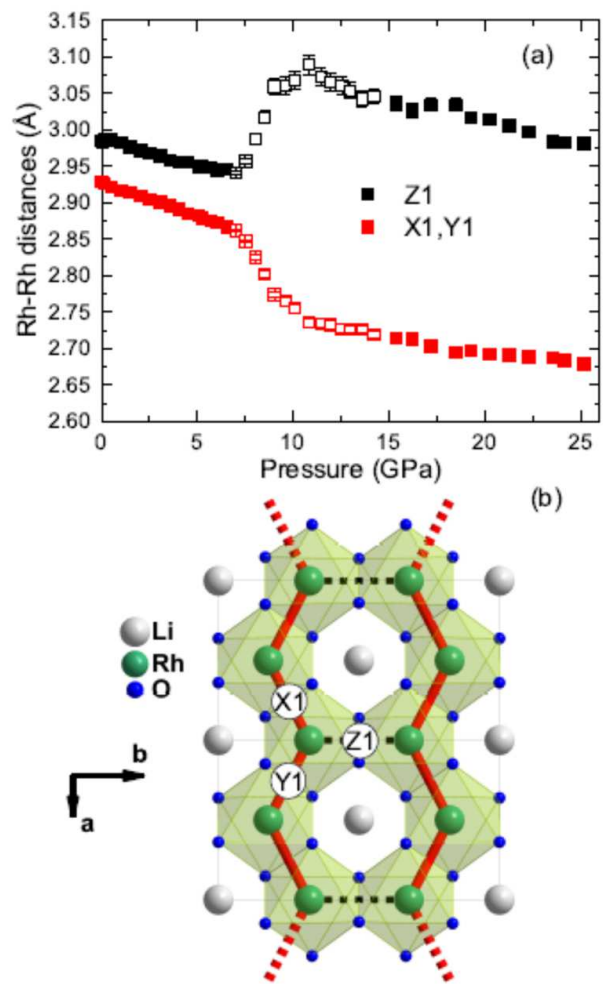}
	\caption{(a) Rh-Rh bond lengths as a function of pressure for the rhodium hexagons in the $ab$ plane with the nomenclature (Rh bonds $X1$, $Y1$, $Z1$) given in (b). The ratio $l_l/l_s$ is calculated to $\approx 1.02$ and $\approx 1.11$ in the low- and high-pressure phases, respectively. The Rh zigzag chains along the $X1$ and $Y1$ bonds above $P_{c2}$ are illustrated in (b) by thick red lines.}\label{fig.Rh-Rh_dist}
\end{figure}

\begin{figure}[t]
	\includegraphics[width=0.85\columnwidth]{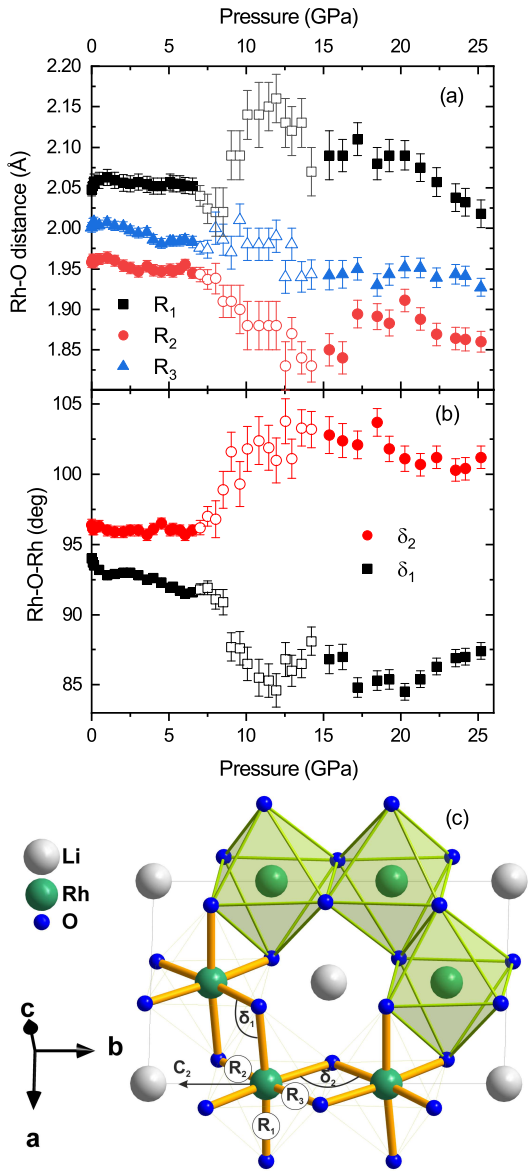}
	\caption{Pressure dependence of the various octahedral (a) Rh-O distances ($R_1$, $R_2$, $R_3$) and (b) Rh-O-Rh bond angles ($\delta_1$, $\delta_2$) with the nomenclature given in (c). The C$_2$ rotational axis is indicated by an arrow.}\label{fig.Rh-O}
\end{figure}

\begin{figure}[b]
	\includegraphics[width=1\columnwidth]{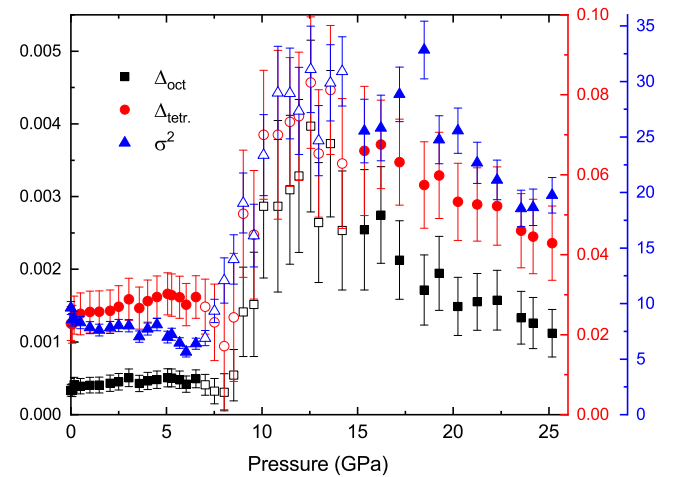}
	\caption{Pressure dependence of the bond-length distortion $\Delta_\text{oct}$, the tetragonal distortion $\Delta_\text{tetr}$, and the bond-angle distortion $\sigma_\text{oct}^2$ as defined in the text.}\label{fig.oct-distortion}
\end{figure}

For a more detailed discussion, the atomic parameters of the refinement are shown in Table \ref{tab.atomic}.  The partial exchange of Li and Rh accounts for the stacking faults, as described in Ref.\ \onlinecite{OMalley.2008}. Since three Rh atoms are required to change place with one Li atom in order to mimic one stacking fault, each 5.5(1) unit cells one stacking fault occurs at ambient pressure. This value is very similar to previous reports in Li$_2$RhO$_3$~\cite{Luo.2013,Todorova.2011} and slightly higher than in $\upalpha$-Li$_2$IrO$_3$ and Li$_2$PtO$_3$~\cite{OMalley.2008}. On the other hand, different studies of Na$_2$IrO$_3$~\cite{Ye.2012,Choi.2012} reported the concentrations of stacking faults that are either larger or smaller than in Li$_2$RhO$_3$.
The number of unit cells per stacking fault increases monotonically with increasing pressure and reaches 7.4(5) at 25.2~GPa, as shown in Fig.~\ref{fig.stackingfaults}, i.e., the number of stacking faults is slightly reduced by external pressure. The parameters for the oxygen positions are changed in the high-pressure phase as compared to the low-pressure phase, thus affecting the RhO$_6$ octahedra. The most interesting change, though, is observed for the $y$-parameter of Rh(1) that determines the Rh-Rh distances in the honeycomb network (see Table \ref{tab.atomic}).

To evaluate this behavior further, we compare the pressure evolution of the three Rh-Rh bond lengths in the $ab$-plane, namely, the $Z1$ bond and the two degenerate $X1/Y1$ bonds as depicted in Fig.\ \ref{fig.Rh-Rh_dist}(b).  At ambient pressure, the $Z1$ bond length amounts to 2.985(3)~{\AA}, while the $X1/Y1$ bond length is 2.9296(13)~{\AA} [see Fig.\ \ref{fig.Rh-Rh_dist}(a)], leading to a slightly distorted honeycomb. The corresponding bond disproportionation $l_l/l_s$, with $l_l$ and $l_s$ being the long and short bonds of the hexagonal Rh network, respectively, amounts to $l_l/l_s$=1.02. For the high-pressure phase, both $X1$ and $Y1$ bonds are drastically reduced by $\approx$0.15~{\AA}, while the $Z1$ bond is increased by the same amount [Fig.\ \ref{fig.Rh-Rh_dist}(a)]. Hence, the bond disproportionation increased to $l_l/l_s$=1.11 at 25~GPa. The $X1/Y1$ bond length of $\approx$2.7~{\AA} above $P_{c2}$ is close to but still larger than the interatomic distances in metallic rhodium ($d$=2.69~{\AA}\cite{Bale.1958}). We thus conclude that external pressure introduces zigzag chains of rhodium atoms along the $a$ direction, as illustrated in Fig.\ \ref{fig.Rh-Rh_dist}(b). A similar structure but with closer bond lengths is found for 5\% Na doped crystals (Li$_{0.95}$Na$_{0.05}$)$_2$RuO$_3$ at ambient pressure \cite{Mehlawat.2017}. On the other hand, pure Li$_2$RuO$_3$ is dimerized at ambient pressure with an armchair pattern of the short Ru--Ru bonds (see Fig.\ \ref{fig1}(e) for illustration) \cite{Mehlawat.2017, Kimber1.2014}.

Next, we consider the pressure-induced changes in the RhO$_6$ octahedra. To this end, we define various Rh--O bond lengths and Rh--O--Rh bond angles that are responsible for the direct metal-to-metal and indirect oxygen-mediated contributions. The octahedra possess a two-fold rotational C$_2$-axis which is indicated by the arrow in Fig.\ \ref{fig.Rh-O}(c). There are three unique Rh--O bonds labeled $R_1$, $R_2$, $R_3$ and two unique Rh--O--Rh angles $\delta_1$ and $\delta_2$, where $\delta_1$ ($\delta_2$) involves two Rh atoms connected via the $X1/Y1$-bond ($Z1$-bond) [see Fig.\ \ref{fig.Rh-O}(c)]. The pressure dependence of the various bonds and bond angles is depicted in Figs.\ \ref{fig.Rh-O}(a) and (b), respectively. At ambient pressure, the largest Rh--O bond length is found for the apical oxygen atom, thus the RhO$_6$ octahedra show a tetragonal distortion with axial elongation. In the low-pressure phase ($P$$<$$P_{c1}$), the bond length $R_1$ is pressure-independent, whereas $R_2$ and $R_3$ slightly decrease under pressure.

At $P_{c2}$, the length $R_1$ is increased as compared to the low-pressure phase, whereas $R_2$ and $R_3$ are decreased. Upon further compression, $R_1$ decreases, $R_3$ seems to be unaffected, and $R_2$ shows a small anomaly at 15-20~GPa that may be significant, as the changes exceed the error bars.

The formation of zigzag chains is predominantly due to a change of Rh-O-Rh angles as described in the following.
The pressure dependence of the Rh--O--Rh bond angles $\delta_1$ and $\delta_2$ is shown in Fig. \ref{fig.Rh-O}(b). At ambient pressure, the values of $\delta_1$ and $\delta_2$ amount to 94.0(3)$^\circ$ and 96.4(4)$^\circ$, respectively. While $\delta_2$ is independent of pressure in the low-pressure phase, $\delta_1$ decreases by increasing pressure. When entering the high-pressure phase above $P_{c2}$ the bond angle $\delta_1$ is strongly decreased to the value 87$^{\circ}$. Interestingly, the onset of the intermediate phase at $P_{c1}$ appears at a pressure, when $\delta_1$ approaches 90$^\circ$, which is a distinct angle for the contributions of the ligand-mediated hopping to the hopping parameters, as discussed in more detail in Refs.~\onlinecite{Winter.2016} and~\onlinecite{Winter.2017}. The strong pressure-induced decrease in the angle $\delta_1$ between $P_{c1}$ and $P_{c2}$ confirms the formation of Rh zig-zag chains along the $a$ direction. Consistently, the bond angle $\delta_2$ is strongly increased, as the $Z_1$ bond length is increased [see Fig.\ \ref{fig.Rh-Rh_dist}(a)]. Again an anomaly is observed for the Rh--O--Rh bond angles between 15-20~GPa, which is directly related to the anomaly for the Rh--O distances and thereby has the same origin.

The electronic states of Li$_2$RhO$_3$ are affected by the distortion of the RhO$_6$ octahedra. Therefore, we followed the pressure dependence of the octahedral distortion using the bond-length distortion $\Delta_\text{oct}$ and the bond-angle distortion $\sigma_\text{oct}^2$~\cite{Ertl.2002,Kim.2004,Hogan.2017}. The bond-length distortion is defined as $\Delta_\text{oct}$=$\frac{1}{6}\sum_{i=1}^{6}[(d_i-d_\text{av})/d_\text{av}]^2$, where $d_i$ is an individual Rh--O bond length and $d_\text{av}$ the average Rh--O bond length in the RhO$_6$ octahedron. The bond-angle distortion is calculated according to $\sigma_\text{oct}^2$=$\frac{1}{11}\sum_{i=1}^{12}(\alpha_i-90)^2$, where $\alpha_i$ is an individual O--Rh--O bond angle. At ambient pressure, the distortion parameters are $\Delta_\text{oct}$=3.4(11)$\times$10$^{-4}$ and $\sigma^2=9.6(6)$ comparable to the results in Ref.\ \cite{Todorova.2011} [$\Delta_\text{oct}$=1.2(6)$\times$10$^{-4}$ and $\sigma^2$=9.6(5)], although the Rh position seems to be fixed in that report. Comparison of our refinement to previous ones~\cite{Mazin.2013,Luo.2013} is not straightforward since in those studies some oxygen parameters were fixed or calculated. The distortion parameters for Li$_2$RhO$_3$ reported in our study are comparable to the ones of the related materials $\alpha$-Li$_2$IrO$_3$ and Li$_2$PtO$_3$ \cite{OMalley.2008,Freund.2016}. For Na$_2$IrO$_3$ the $\Delta_\text{oct}$ value is about one magnitude smaller, while the bond-angle distortion $\sigma^2$ is nearly doubled \cite{Ye.2012,Choi.2012}. A comparison to the octahedral distortions in dimerized Li$_2$RuO$_3$ is difficult, since the reported values determined by various studies are not consistent. For example, the $\Delta_\text{oct}$ values between 1.4$\times$10$^{-4}$ and 24$\times$10$^{-4}$ have been reported, and the values for $\sigma^2$ range between 4.7 and 54 \cite{Wang.2014,Kobayashi.1995,Mehlawat.2017}.

In the low-pressure phase, the bond-length distortion only slightly increases with increasing pressure, whereas the bond-angle distortion decreases (Fig.\ \ref{fig.oct-distortion}). At the critical pressure $P_{c2}$, both parameters $\Delta_\text{oct}$ and $\sigma_\text{oct}^2$ are drastically enhanced compared to the low-pressure range. Such an enhanced distortion was also reported in Ref.\ \cite{Wang.2014} for dimerized Li$_2$RuO$_3$ compared to the non-dimerized samples. It is therefore likely that the enhancement of $\Delta_\text{oct}$ and $\sigma_\text{oct}^2$ at $P_{c2}$ is caused by the lattice strain due to the formation of the Rh--Rh zigzag chains in Li$_2$RhO$_3$.

Of further interest is the tetragonal distortion (elongation or compression along the $z$-direction) of the octahedra, as this would cause a splitting of the Rh $t_{2g}$ states. As a measure of the tetragonal distortion we define the parameter $\Delta_\text{tetr}$ as the deviation of the apical Rh--O bond length $R_1$ from the average Rh--O bond length $d_\text{av}$ according to $\Delta_\text{tetr}=(R_1-d_\text{av})/d_\text{av}$ \cite{Kim.2004}. For positive (negative) nonzero values of $\Delta_\text{tetr}$ the octahedra are elongated (compressed) along the apical bond direction. Such a distortion can be explained by a cooperative first-order Jahn-Teller effect neglecting stress on the system \cite{Gehring.1975,Kaplan.1995}. The Jahn-Teller effect is expected to be weak but nonzero in a $d^5$ configuration. The pressure dependence of $\Delta_\text{tetr}$ is depicted in Fig.\ \ref{fig.oct-distortion}. We note that the tetragonal distortion at ambient pressure amounts to $\Delta_\text{tetr}$=0.023(4), which is comparable to the value $\Delta_\text{tetr}$=0.015(3) given in Ref.\ \cite{Todorova.2011}. In the low-pressure phase, $\Delta_\text{tetr}$ increases slightly but steadily upon compression, i.e., the elongation increases.

Between the low- and high-pressure phases, $\Delta_\text{tetr}$ is nearly doubled, before it decreases upon further compression above $P_{c2}$. While the tetragonal distortion in the low-pressure phase is comparable to that in Li$_2M$O$_3$ ($M$= Ir, Pt), it is much more pronounced than in Na$_2$IrO$_3$, where $\Delta_\text{tetr}$ is close to zero. We thus conclude that the lattice strain rather than the Jahn-Teller effect is the driving force for the distortion of the octahedra in the honeycomb lattices. The tetragonal distortion in the high-pressure phase of Li$_2$RhO$_3$ is comparable to the tetragonal distortion in the perovskites Sr$_2$RhO$_4$ and Sr$_2$RuO$_4$, where the bond angle distortion is zero \cite{Huang.1994,Vogt.1996}.

\subsection{Computational results}

A question that remains open is why \lirho\ retains the monoclinic $C2/m$ symmetry and shows the zigzag-chain pattern of short Rh--Rh bonds under pressure, whereas \liiro~\cite{Hermann.2018,Clancy.2018} and $\alpha$-RuCl$_3$~\cite{Bastien.2018} become triclinic following the formation of metal-metal dimers.

In previous studies, the experimentally observed dimerization pattern in \liiro\ and $\alpha$-RuCl$_3$ was identified by DFT calculations within the GGA+SOC+$U$ scheme~\cite{Hermann.2018}, as a consequence of a complex interplay of SOC, magnetism, correlation, and covalent bonding. Following this knowledge, we performed first full relaxations of \lirho\ as a function of hydrostatic pressure with and without SOC. As initial guess for the geometrical optimization at each pressure, we considered two structures. The experimental low-pressure `undistorted' structure at 5 GPa and the experimental high-pressure `zigzag chain' structure at 25.2 GPa. In our notation, we assume a structure to be undistorted when the corresponding bond disproportionation $l_l/l_s < 1.04$. Moreover, for each of these initial geometries, we considered five different spin configurations, as explained in Section~\ref{sec:method-comput}.

\begin{figure}[!t]
	\includegraphics[width=0.9\columnwidth]{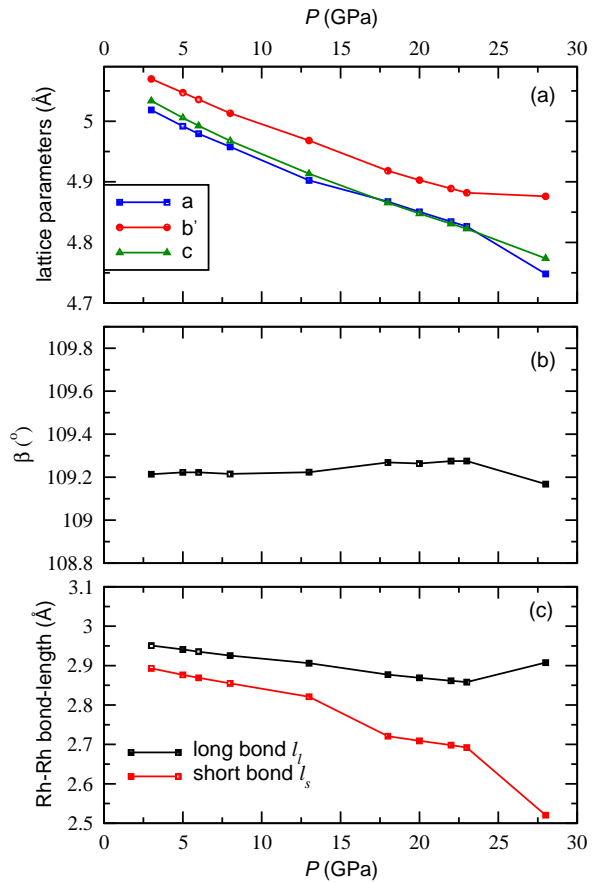}
	\caption{Pressure dependence of the theoretically obtained (a)-(b) structural parameters calculated under hydrostatic pressure conditions and (c) Rh-Rh bond lengths within the GGA+SOC+$U$ ($U=1.5$ eV) scheme. }
\label{fig.theo3}
\end{figure}

Test calculations performed at 25 GPa reveal that after relaxation the structure becomes dimerized, regardless of the initial configuration.
At a given pressure, the energetics of the various different configurations are obtained by comparing the corresponding enthalpies. Due to the underbinding problem of GGA (relaxed interatomic distances are longer than their experimental counterparts), the volume corresponding to 2 GPa reproduces the experimental volume at ambient pressure. This has been corrected by systematically subtracting $\Delta P=2$~GPa from all simulated pressure values.

The value of Hubbard correlation $U=1.5$ eV was chosen such that at 5 GPa (within GGA+SOC+$U$ scheme): (a) the optimized lowest-enthalpy magnetic configuration corresponds to $C2/m$ symmetry and reproduces the experimental value of $l_l/l_s$ = 1.02, and (b) the nonmagnetic configuration is dimerized (though parallel $X1/Y1$ type). The latter confirms dimerization at finite pressure for  \lirho\  as magnetism is known to work against dimerization by pushing the transition pressure to a higher value~\cite{Hermann.2018}.

At a pressure $P_c ^{\rm noSOC}$$\approx$11 GPa, we find that \lirho\ undergoes a phase transition from a homogeneous to a dimerized phase with bond disproportionation $l_l/l_s $=1.146 within the GGA+$U$ scheme (not shown here). Upon dimerization, \lirho\ becomes nonmagnetic.
However, there are a few discrepancies with the experimental structures: (i) below $P_c^{\rm noSOC} $, in the homogeneous structure, the shorter bond corresponds to the $Z1$ bond, rather than to the $X1$ and $Y1$ bonds as observed in experiments, and (ii) the dimerized phase does not have the $C2/m$ symmetry, rather it has the triclinic (\p1bar) symmetry, similar to \rucl{}~\cite{Bastien.2018} and \liiro~\cite{Hermann.2018,Clancy.2018}. The inclusion of SOC reproduces the shorter bond to be  $X1/Y1$ in the homogeneous phase (below  $P_c^{\rm SOC} $) and shifts the transition pressure $P{_{cm}}$ to 27 GPa. However, the high-pressure structure still becomes triclinic with $l_l/l_s $=1.15 (see Fig.\ \ref{fig.theo3}).

\begin{figure}[t]
	\includegraphics[width=1\columnwidth]{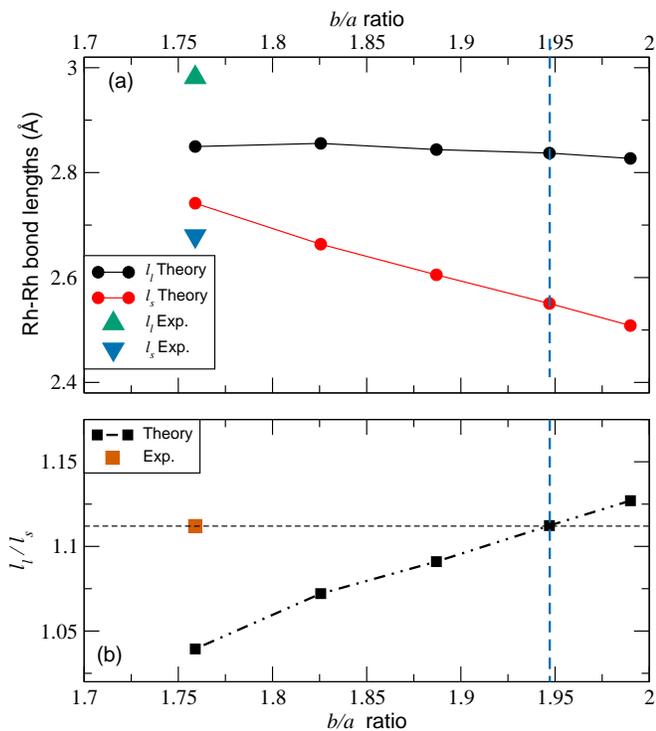}
	\caption{Variation of (a) the Rh-Rh bond lengths  and (b) bond-disproportionation ($l_l/l_s$)  as a function of the $b/a$ ratio, with the $b$ and $c$ parameters fixed to its experimental value at 25.2 GPa (within GGA+SOC+$U$), and comparison with experimentally obtained values. The dashed blue lines show the optimal value of $b/a$ ratio that illustrates the choice of lattice parameters for uniaxial pressure conditions in the simulation. }
\label{fig:uni}
\end{figure}

\begin{figure}[t]
      \includegraphics[width=1.3\columnwidth]{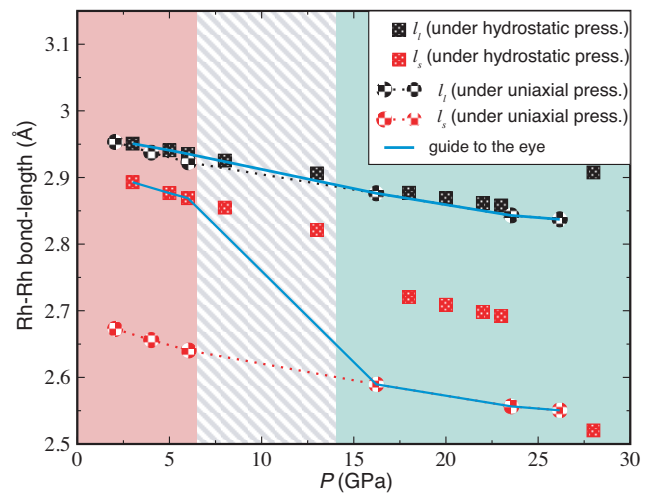}
\caption{The evolution of  Rh-Rh bond lengths (calculated within GGA+SOC+$U$)  as a function of pressure. Red and green shaded regions represent hydrostatic and uniaxial pressure regimes, respectively. For comparison with the experimental data, the blue line is drawn as a guide to follow the transition from hydrostatic to uniaxial pressure. }
\label{fig:eff}
\end{figure}

The above results show that the experimental zigzag pattern of the short Rh--Rh bonds cannot be obtained from hydrostatic pressure simulations. We therefore proceed by simulating uniaxial pressure with the $b$ and $c$ parameters fixed to their experimental values at 25.5 GPa and the $b/a$ ratio varied systematically (Fig.~\ref{fig:uni}). This approach yields the zigzag-chain structure observed experimentally. However, we had to increase the $b/a$ ratio to 1.95, in order to reproduce the ratio $l_l/l_s$ between the long and short bonds.

For obtaining the pressure evolution of Rh-Rh bonds under uniaxial condition, we next repeated the above calculations but this time with $b/a=$ 1.95, while fixing the $b$ and $c$ parameters to their experimental values at the corresponding pressure. By comparing the results of the hydrostatic and uniaxial pressure simulations, we conclude (see Fig.~\ref{fig:eff}) that the evolution of Li$_2$RhO$_3$ up to $P_{c1}$ is compatible with hydrostatic pressure conditions, whereas at higher pressures the system progressively moves toward the behavior expected under uniaxial pressure. The uniaxial pressure accounts for the formation of zigzag chains instead of dimers, although it does not fully account for the evolution of the longer Rh--Rh bonds that evolve smoothly in the simulation but show a step-like anomaly experimentally (Fig.~\ref{fig.Rh-Rh_dist}).

\begin{figure}[!t]
    \includegraphics[width=1\columnwidth]{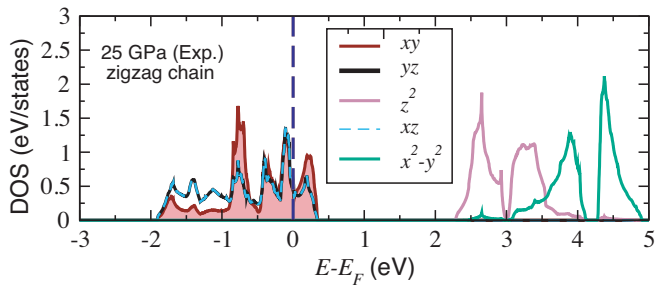}
\caption{Orbital-projected DOS for the Rh $d$-orbitals in the experimental structure with $C2/m$ symmetry at 25.5 GPa, calculated within GGA+SOC+$U$ scheme with $U$=1.5 eV. }
\label{figdos}
\end{figure}

At ambient condition, \lirho\  is an insulator as shown in Ref.~\onlinecite{Mazin.2013}. Our calculated density of states (DOS) for the experimentally obtained structures at 25.2 GPa (Fig.~\ref{figdos}) show that unlike other dimerized phases in  \liiro\ and $\alpha$-RuCl$_3$, in \lirho\ the degeneracy between $yz$ and $xz$ orbitals of Rh $d$ states does not get lifted as the symmetry remains the same. Moreover, the system probably becomes metallic under pressure due to the formation of zigzag chains, which provide new hopping pathways.

The origin of the uniaxial-like pressure conditions requires further investigation. Experimental pressure conditions in a DAC with helium as pressure-transmitting medium are expected to be hydrostatic. Therefore, we consider the nature of the Li$_2$RhO$_3$ sample as a more plausible reason. In particular, stacking faults that occur, on average, at every 6-7 layers could act as a local strain and affect the evolution of the structure under pressure. Our data show that the concentration of stacking faults in Li$_2$RhO$_3$ is higher than in the polycrystalline samples of $\alpha$-Li$_2$IrO$_3$ and in single crystals of $\alpha$-Li$_2$IrO$_3$ that were used in our previous study~\cite{Hermann.2018}. Interestingly, Na$_2$IrO$_3$ shows different pressure evolution of the crystal structure in powders~\cite{Xi.2018} and single crystals~\cite{Hermann.2017}. Given the proclivity of Na$_2$IrO$_3$ to the formation of stacking faults, a similar mechanism may be operative there and deserves further systematic investigation.

\section{Conclusion}
In contrast to $\alpha$-Li$_2$IrO$_3$ and $\alpha$-RuCl$_3$, where a dimerized triclinic phase is stabilized under pressure, Li$_2$RhO$_3$ retains its ambient-pressure monoclinic symmetry and develops zigzag chains of short Rh--Rh bonds. This structural phase transition is not abrupt, since traces of the low-pressure phase can still be found even at the highest pressure of 25.2~GPa, but above $\approx$14~GPa the high-pressure phase is dominant. Our density-functional calculations suggest that such a behavior is not anticipated in Li$_2$RhO$_3$ under hydrostatic pressure, where conventional dimerization should occur. On the other hand, uniaxial pressure may explain the experimental observations and promote the formation of zigzag chains instead of dimers.

\vspace*{1em}
\begin{acknowledgments}
We thank the ESRF, Grenoble, for the provision of beamtime at ID15B. DKh, RV, AAT, and PG acknowledge financial support by the Deutsche Forschungsgemeinschaft (DFG), Germany, through TRR 80, SPP 1666, TRR 49, and SFB 1238. AJ acknowledges support from the DFG through Grant No. JE 748/1. SB thanks Stephen Winter for helpful discussions. AAT acknowledges financial support from the Federal Ministry for Education and Research via the Sofja-Kovalevskaya Award of Alexander von Humboldt Foundation, Germany.
\end{acknowledgments}

%

\end{document}